\documentclass[10pt,twocolumn]{article}
\usepackage{graphicx}
\usepackage[normalem]{ulem}
\usepackage{amsmath}
\usepackage{amsfonts}
\usepackage{cite}
\usepackage{upgreek}

\newcommand{\MYcomment}[1]{}

\newcommand{\MYnote}[1]{}

\newcounter{MYtablecntr}
\addtocounter{MYtablecntr}{1}

\newcommand{\MYlabel}{\small {$\bullet$}}

\newcounter{MYenumctrtwo}

\newcounter{MYenumctr}

\usepackage{algorithm}
\usepackage[noend]{algpseudocode}

\title{Early Work on Efficient Patching for Coordinating \\
Edge Applications 
}

\author{ \large
    Naveen T.R. Babu and Christopher Stewart\\
         Department of Computer Science and  Engineering, The Ohio State University\\
}

\begin{document}
%\author{Naveen Tumkur Ramesh Babu}
%\affiliation{%
%  \institution{The Ohio State University}
%  \streetaddress{P.O. Box 1212}
%  \city{Columbus}
% \state{Ohio}
% \postcode{43201-6221}
%}
%\email{naveentumkurrameshbabu.1@osu.edu}

\date{}
\maketitle
\thispagestyle{empty}
\pagestyle{empty}

%%%%%%%%%%%%%%%%%%%%%%%%%%%%%%%%%%%%%%%%%%%%%%%%%%%%%%%%%%%%%%%%%%%%%%%%%%%%%%%%
\begin{abstract}
%%% Things to talk about in abstract:-
%  Current AVs and where research is heading
% 1. What is the new technology (DONE)
% 2. Why is it important? (DONE)
% 3. What is the problem? (DONE)
% 4. How much does the problem inhibit technology (DONE)
% 5. What is the expected type of solution? (DONE)
% 6  Why is it hard to achieve? (DONE)
% 7. What is your solution? (DONE)
% 8. How does your solution get around the hard problem in (quesn6)
% 9. What will I do (Since this is an idea paper)
% 10. Why will all of this matter?
Multiple applications running on Edge computers can be orchestrated to achieve the desired goal. Orchestration of  applications is prominent when working with Internet of Things based applications, Autonomous driving and Autonomous Aerial vehicles. As the applications receive modified classifiers/code, there will be multiple applications that need to be updated. If all the classifiers are synchronously updated there would be increased throughput and bandwidth degradation. On the other hand, delaying updates of applications which need immediate update hinders performance and delays progress towards end goal. The updates of applications should be prioritized and updates should happen according to this priority. This paper explores the setup and benchmarks to understand the impact of updates when multiple applications working to achieve same objective are orchestrated with prioritized updates. We discuss methods to build a distributed, reliable and scalable system called "DSOC"(Docker Swarm Orchestration Component).

%--  Think of this like a Micro service architecture, each component logically separate and performing their functionality. When we have such Docker containerized application running and they need model updates, there should be an efficient way to track each application and prioritize which application model need to be updated. There should be an engine which tracks the performance of system and decide which application need to be updated with a newer model.
\end{abstract}

%%%%%%%%%%%%%%%%%%%%%%%%%%%%%%%%%%%%%%%%%%%%%%%%%%%%%%%%%%%%%%%%%%%%%%%%%%%%%%%%

\section{INTRODUCTION}

%%% Things to talk about in introduction:-
%  Current AVs and where research is heading
% 1. What is the new technology (DONE)
% 2. Why is it important? (DONE)
% 3. What is the problem? (DONE)
% 4. How much does the problem inhibit technology (DONE)

% Make sure the above points are strong enough in abstract and expand each point in the introduction.....

% Draw a graph explaining how current trend of application dependence on smaller components lead to end goal.
% 1. How many applications coordinate together towards end goal
% 2. How many such co-ordinating application have updates?
% Prioritize the updates: Critically important, major update, minor update and unimportant --- Green, Yellow, blue and Red

Autonomous systems like Self-driving cars, autonomous aerial systems, smart restaurants and smart traffic lights have attracted a lot of interest from both academia and Industry~\cite{transportresearch}~\cite{boubin2019managingdeprecated}. Many top companies like Google, Uber, Intel, Apple, Tesla, Amazon have significantly invested in researching and building Autonomous Systems~\cite{selfdriving}. An autonomous system is a critical decision making system which makes decisions without human intervention. An autonomous system is comprised of complex technologies which learns the environment, makes decision and accomplishes the goal~\cite{transportresearch}. In this paper, we focus on "Microservices Orchestration" for coordinating multiple autonomous applications that are working towards a common objective.
 
Microservices is an architectural style which structures an application as a collection of different services that are loosely coupled, independently deployable and highly maintainable. Large, complex applications can be deployed using Microservices where each service will have different logical function and contribute to the bigger application~\cite{micro_article}. %There is high Complexity coordinating Microservices based application~\cite{micro_article}.
When working towards a particular goal, we might need to deploy multiple applications which need to take up a sub task and coordinate with other applications in order to efficiently complete the task at hand. Efficient coordination of multiple different applications is really crucial for building fully autonomous systems.% Handling interaction between applications, error handling and failure recovery is resilient. 
Configuring, controlling and keeping track of each microservice would be really hard~\cite{orchestrate}. An efficient way to track and manage multiple applications would be using an orchestrator.  Orchestration is a process which involves automated configuration, management and coordination of multiple computer systems and application software~\cite{it_orchestrate}. There are various orchestration tools for Microservices like ansible~\cite{ansible}, Kubernetes~\cite{kubernetes}, Docker Swarm~\cite{swarm_docker}. 

%############# Insert one citation above and search for different types of orchestrator. ############
%####### add 2-3 sentences also

%%%%%%%%%% Please insert graph1 here.
\begin{figure}
    \centering
    \includegraphics[width=8cm]{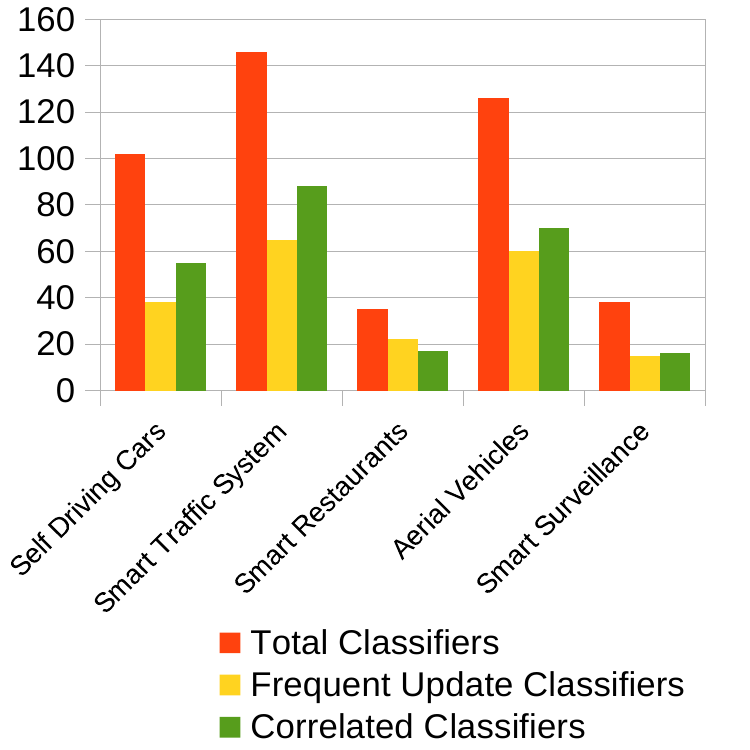}
    \caption{Why Coordination and Patching?}
    \label{fig:introduc}
\end{figure}

When working with Artificial Intelligence based applications, the performance of each microservice may degrade over time and there needs to be an updated code or Machine Learning model in order to restore the performance~\cite{naveeniotonline}. When multiple applications seek an update, allowing all updates would degrade bandwidth, increase throughput and may not yield much performance gain~\cite{naveeniotonline}. If any microservice application need an update, it would be a tedious task to identify individual application and perform the update. While performing such updates we need to consider individual application performance, progress towards end goal and system performance~\cite{7886112}. It is practically impossible to consider every application's performance parameters and pick the model to be updated at run-time~\cite{poster_Aiiot}.

A Patch could be a code update that fixes a bug or yields performance improvement, Machine Learning model update and would be referred as "Classifier". The term Classifier is used in rest of this paper. Figure~\ref{fig:introduc} approximates the usage of Classifiers in AI-based applications and patching in real world Autonomous Systems such as Self Driving cars, Smart Traffic Systems, Aerial vehicles and Smart surveillance. These autonomous applications make use of nearly 40-140 total classifiers out of which at least 40 percent have frequent classifier update to improve performance~\cite{naveeniotonline}. The frequency of update in individual application is calculated by performing a literature survey of updates using incremental software releases~\cite{autopilot}~\cite{dji}~\cite{trlights}.
%%%%% ############ Try to add a couple of citations above.
Out of the total updates, at least 50 percent of them are correlated updates. For example, an update to an application's model would impact the performance of another interdependent model or code fragment. If multiple applications are coordinating with one another towards a common objective, the choice of update significantly impacts the performance of the system and rate of completion towards the end goal.
% The performance can be measured in terms of accuracy, latency and execution time of the application. 

This paper proposes Docker Swarm Orchestration Component called "DSOC" which is responsible for orchestrating multiple applications and efficiently prioritizing classifier updates. To the best of our knowledge, this is the first work to propose an efficient method of using Docker Swarm for multiple AI based application coordination involving classifier updates. 
%The expected type of solution would be to regularly store the performance metrics of individual microservice application. When multiple applications seek an update, a program logic can be run to choose to rank the applications based on importance of update. This would incur a lot of processing at the time of performing update and there might be deadlock situation when only few updates can be performed synchronously but many applications are striving for immediate update [5].
 %An orchestrator would track and record the status and performance of multiple coordinating applications. The orchestrator would take care of tracking different applications and take complete responsibility of seamlessly updating the applications.

% DONEEEEE --- Inserting figure is causing all sorts of alignment issue.
%****** Draw a graph explaining what the current trend is and how an orchestrator like K8 or Docker swarm would help while running dynamic application which are working towards the same objective.
%********

\section{Background}

\subsection{Docker and Docker Swarm}
Docker Containers are best suited for Microservices. Docker provides lightweight encapsulation of each application enabling independent deployment and scaling of each microservice. Docker is a composing Engine for Linux containers, an OS-level virtualization technique, which uses namespace and cgroups to isolate applications in the same linux kernel. Control group (abbreviated as cgroup) is a collection of processes that are associated with a set of parameters. cgroup ensures that the specified resources are actually available for a container. Namespace isolation is another such feature where groups of processes are separated such that each group cannot see the resources used by other groups. The kernel resources are partitioned such that a set of processes sees a set of resources while another set of processes see a different set of resources~\cite{docker_article}. 

Docker uses copy on write (COW) and layered storage within images and containers. A Docker image is a read-only template, which references a list of read-only storage layers, used to build a linux container. Docker container is a standard software unit that packages up code and dependencies so that the application runs quickly and can be shipped reliably from one computing environment to another. Docker images become Docker containers at run time when they run on Docker Engine. The layered storage allows fast packaging and shipping of an application as a lightweight container by sharing common layers between images and container. By using Docker, there is potential for faster deployment time and faster model updates~\cite{docker_article}.

When you have a lot of containerized applications running, there should be a mechanism to make them all work towards a common goal. One method to achieve this is using Docker Swarm. Docker Swarm is a group of machines that are joined together as a cluster and commands are executed by swarm manager to control the group of machines. Each machine in swarm is called a node which can be physical or virtual Machine. Applications can be defined using a manifest file and easily deployed using Docker commands~\cite{swarm_docker}.

\subsection{Worker and Manager Nodes}
Manager Nodes control the cluster with tasks such as maintaining the state of the cluster, dispatching tasks to worker and providing fault tolerance to the system. Currently, Docker supports using multiple Manager nodes where only one manager would be elected as leader and it performs all the responsibilities of a manager. The other manager nodes are standby managers which receive updated information about state of the system and may be chosen as leader when leader node goes down. Using multiple managers is fairly new experimental feature in Docker which would be explored in this research. Worker Nodes are instances which accept tasks from Manager Nodes and execute them as containers. Worker Nodes will not share their state information with other worker nodes and do not make scheduling decisions~\cite{swarm_docker}.

\subsection{Scale-in and Scale-out of applications}
When a Microservice application is deployed, we might need to increase the number of microservice components (scale-out) or decrease the number of microservice components (scale-in) based on the user demand and progress towards end goal. This calculation should happen automatically and applications need to be re-scaled based on workload and progress towards end goal~\cite{Venugopal_2017}.

\begin{figure}
    \centering
    \includegraphics[width=8.5cm]{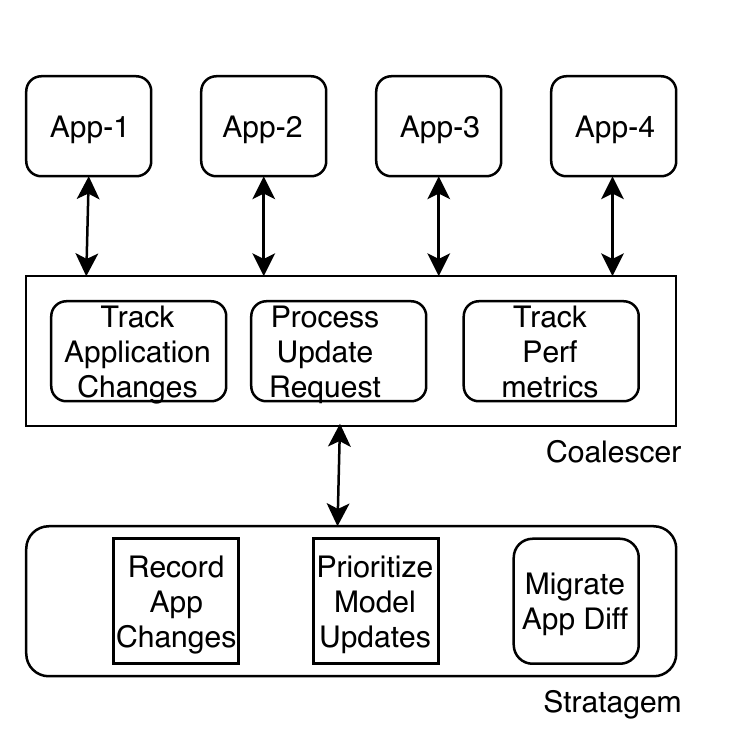}
    \caption{System Architecture diagram}
    \label{Design2}
\end{figure}

\section{System Architecture}
As depicted in Figure~\ref{Design2}, the System architecture consists of 3 main components: Application, Coalescer and Stratagem. Applications are lightweight and containerized units which are deployed to achieve a particular sub-task. The main focus in this research is to choose updates efficiently when a group of different applications coordinate to achieve a common objective. Such a group of applications working towards a common objective is called "Swarm"~\cite{swarm_docker}.

The single point of contact for multiple applications is the "Coalescer". Coalescer is the orchestrator unit in our design which helps in coordinating multiple applications to achieve a particular goal. A Coalescer has multiple functionalities: It tracks application changes, it processes migration request, it tracks progress and performance per application. If there is performance degradation in any of the application(s), Coalescer makes sure the expected performance of application would be restored. Coalescer handles coordination of multiple applications, updates to application and makes sure overall performance of the system is preserved. Stratagem is a component which records application changes and updates the application with suitable code/model in order to satisfy performance criteria. Stratagem prioritizes updates considering different performance metrics and migrates the required difference, between source code/model and updated code/model, to the appropriate application.

\begin{figure}
    \centering
    \includegraphics[width=8.5cm]{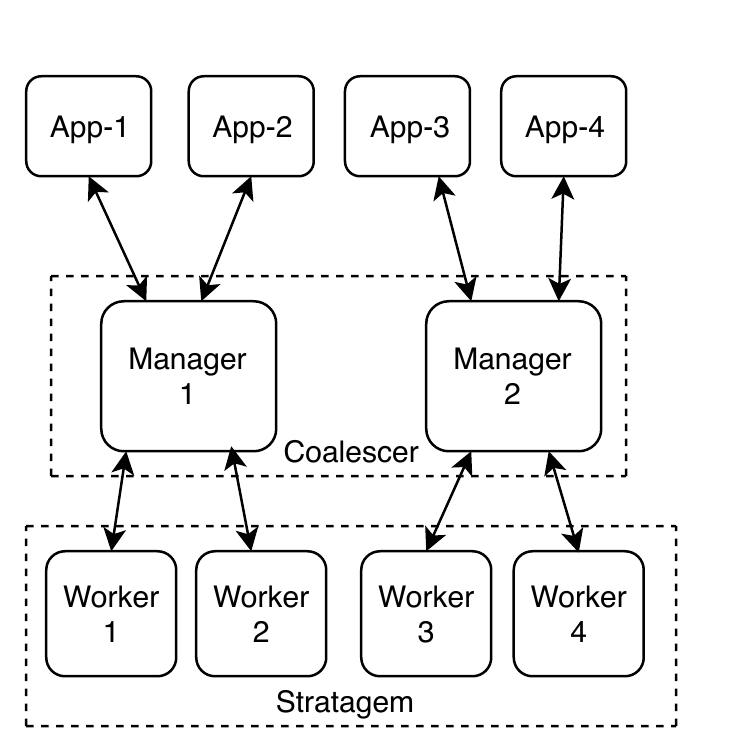}
    \caption{Interaction of different nodes in DSOC}
    \label{figX}
\end{figure}
Figure~\ref{figX} explains the logical components that are used to build Coalescer and Stratagem. "Manager" is a logical component present in Coalescer. A Manager node tracks one or more applications and makes sure the performance of those applications are optimum. "Worker" is a logical component present in Stratagem. A Manager node creates multiple worker nodes to track the deployed applications. A manager creates a worker node per application to track changes and makes sure the performance, progress expectation of that application is being met over time.

As Autonomous systems such as Self driving cars, Aerial Vehicles, Smart traffic system, smart restaurants are becoming increasingly popular~\cite{swarm_docker}, they have not focused on building DSOC type application which efficiently progresses towards the goal, using a strategy which is easy to deploy and maintain. An autonomous application which is deployed in production will comprise of several smaller applications coordinating together to achieve an end goal~\cite{IOT_growth}. This is a Microservices based architecture where each independent component would have a logical function and contributes towards the end goal~\cite{micro_article}. Building such an efficient system which tracks and makes timely progress towards end goal is really crucial. During deployment, there might be updates to individual applications which improves their performance. If all the update hungry applications are allowed to update their model, it would lead to increased throughput and bandwidth degradation. There should be an effective method of prioritizing updates taking several factors such as latency, progress, cpu utilization, memory, accuracy when multiple applications are seeking an update. The implementation section discusses details about prioritizing updates and using the framework from ~\cite{SoftwarePilot}. Using a DSOC approach would give greater control over applications and ensure performance of the system is maintained. Using DSOC, critical concerns like code update, Machine learning model update, performance based progress towards end goal are carefully considered.

\section{Implementation}
%--- Flow chart for basic call flow

%--- Algorithm to prioritize app updates and update process

% The algorithm to prioritize application's model updates would consider
% 1) System Specific parameters: CPU utilization, tput, storage and bandwidth
% 2) Application Specific parameters: Accuracy, Latency, execution time, progress towards end goal, Number of currently processing jobs
\begin{figure}
    \centering
    \includegraphics[width=8cm]{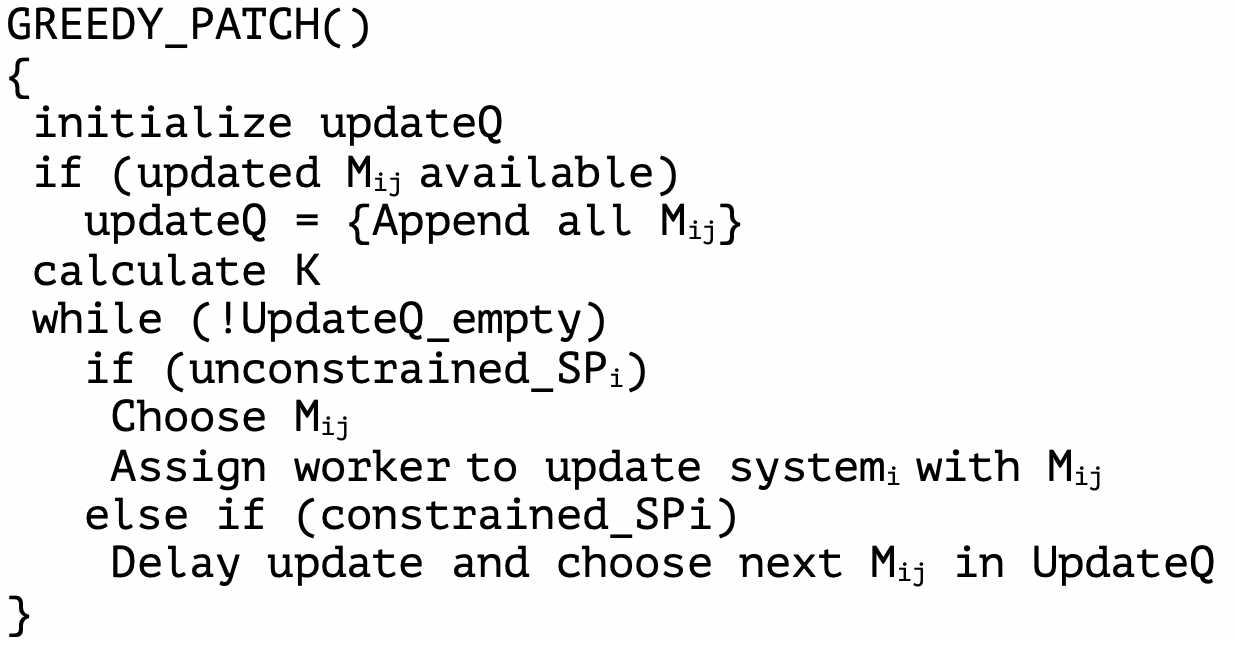}
    \caption{Greedy Algorithm}
    \label{fig:greedy}
\end{figure}

\begin{figure}
    \centering
    \includegraphics[width=8cm]{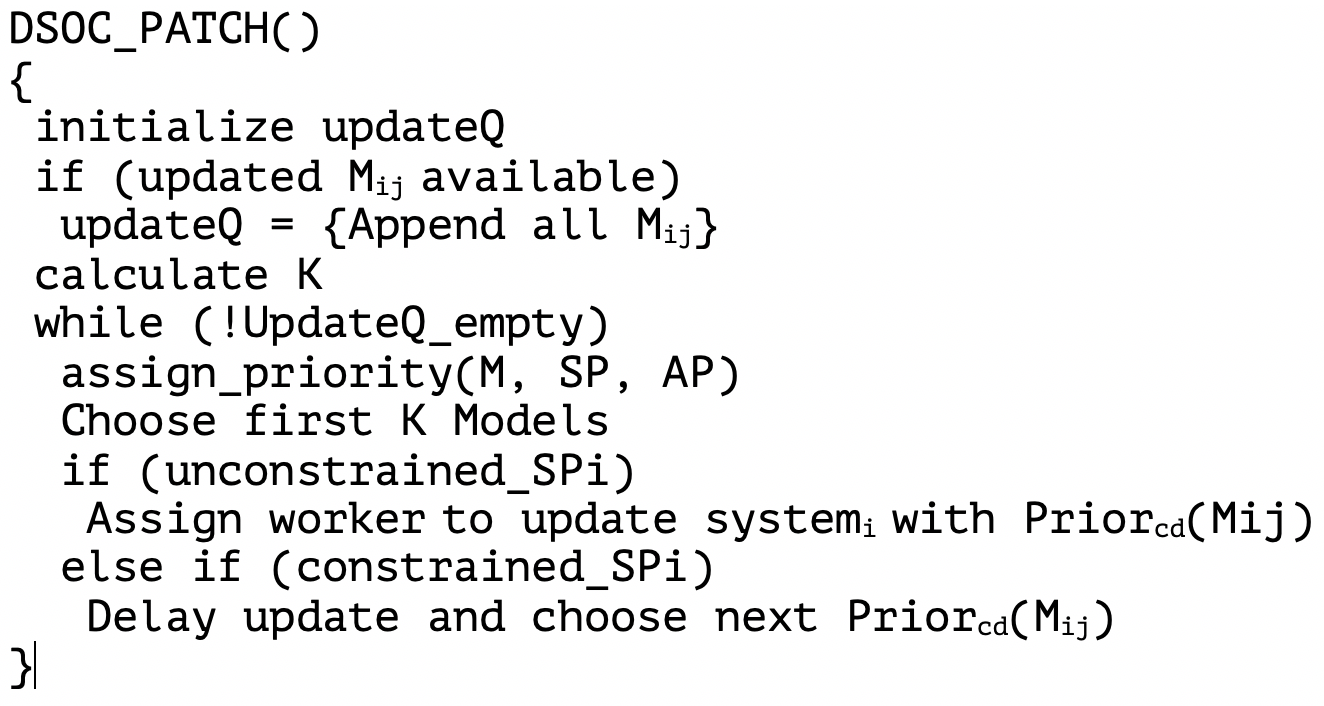}
    \caption{DSOC Algorithm}
    \label{fig:dsoc}
\end{figure}

% We will use the below type of prioritization...
% Overall 0>prior_val>0.3 --- Green
% 0.31 - 0.6 --- Yellow
% 0.61 - 0.85 ---- Blue
% >0.85 --- Red
% Idea --- 1. Use System parameters and 2. Application paramaters
% * Set priority for System parameter and application parameter sum(alphai) =1

% Individual weights for System specific parameters
% * CPu utilization, tput, memory, storage a1,a2,a3 etc
% System performance update() left to individual System

% Individual weights for the application specific parameters
% * Accuracy, individual progress (expected vs current progress) at time-i, latency, 
% Application performance meterics update() left to the individual application

%***** AM I NOT thinking about the whole system performance?????
% ***** Make a claim that If individual apps add up their performance accordingly by making model updates considering system specific parameters and application specific parameters, we should be able to improve the overall system performance.

%%%% Important noteeee ----- Think about simulation, instead of actual deployment on the field....
%*** These new things deployed on field would lead to devastation if things go wrong. Simulation would be the best to check. How do we guarantee the performance we get out of simulation on field? *****
We can leverage the existing Docker swarm functionality for our implementation. Swarm managers control all the nodes and they can use several strategies to run containers efficiently. It can be 1. "emptiest node" technique - which fills the least utilized machine with containers 2. "global"- which ensures each machines gets exactly one instance of the specified machines. These strategies help in load balancing, scaling and fault management~\cite{swarm_docker}.
There are two methods to implement coordination among groups of applications working towards a specific goal. Greedy approach is one method where every application is eager to increase it's accuracy and performance. Whenever there's a newer model or updated code available which improves accuracy and performance, the application tries to perform an update. Figure~\ref{fig:greedy} explains how a greedy approach for patching works. We maintain an update Queue which stores all the model and code updates of applications. M$_{ij}$ is a code/model update for application-j running on node-i. Calculate 'K' updates which can be performed such that overall system performance doesn't degrade. Till the update Queue is non-empty, choose an update M$_{ij}$ and check if the node-i is unconstrained. If it's unconstrained, assign a worker to update the application-j on node-i with M${ij}$. If the node-i was constrained, delay the M$_{ij}$ update and proceed by choosing the next model in update Queue.

The second approach is the DSOC approach (Figure~\ref{fig:dsoc}) where the "Coalescer" handles all the updates, evaluating the priority of update requests. The system specific parameters like throughput, memory, cpu utilization, bandwidth and application specific parameters like accuracy improvement, execution time, latency are carefully considered before updating an existing model/code fragment.

\begin{figure}
    \centering
    \includegraphics[width=8.5cm]{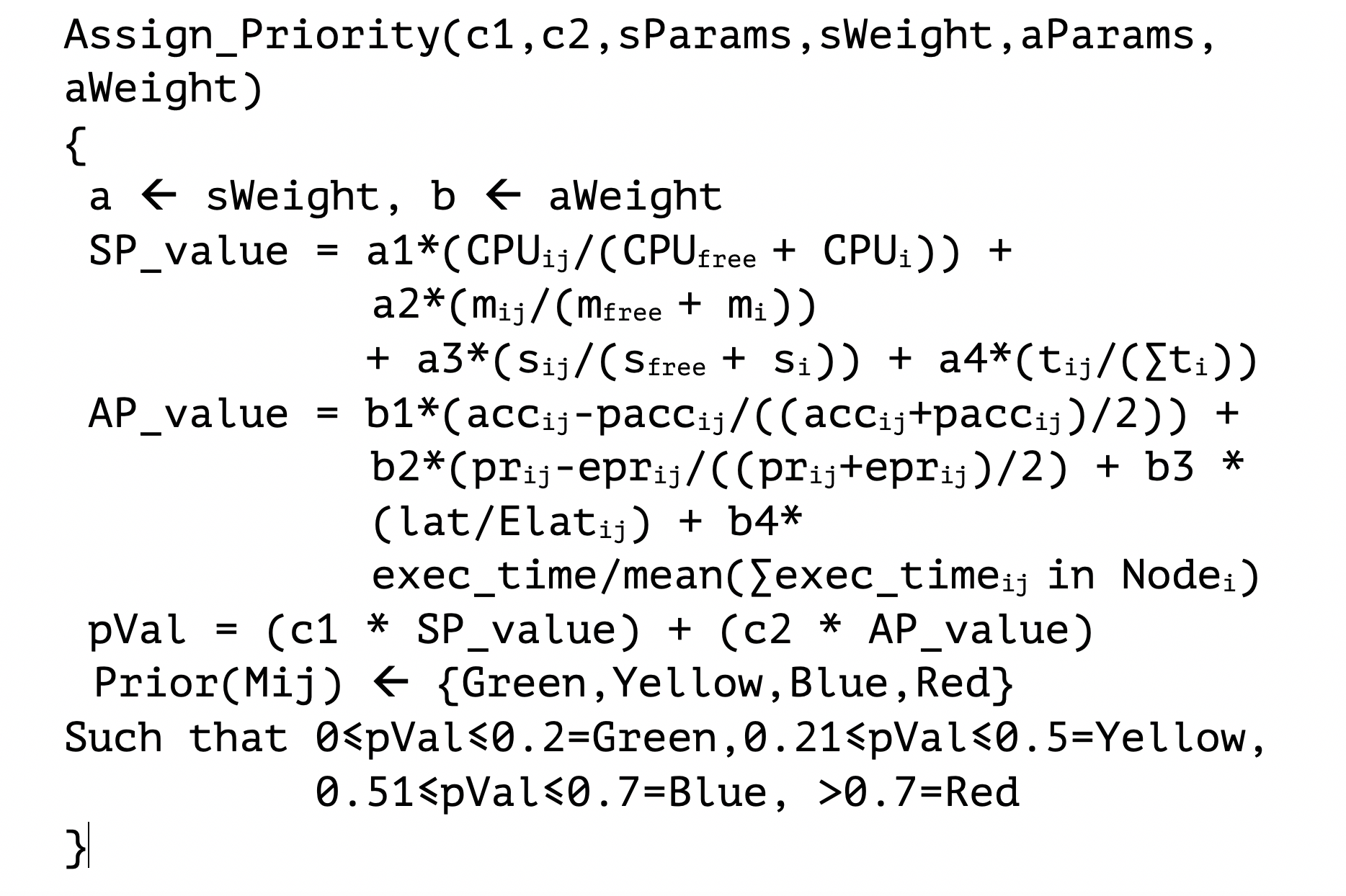}
    \caption{assign\_priority calculation}
    \label{fig:assprior}
\end{figure}

Refer to Figure~\ref{fig:assprior} to understand how priority is assigned to model/code update. The updates are prioritized after considering all system specific and application specific parameters. c1, c2 are the parameters used to indicate the weight to be given to SP(system specific parameters) and AP(Application specific parameters) such that $0\le{c1} \le {c2} \le {1}$ and c1 + c2 = 1. 
System specific weights for CPU utilization, memory, storage and throughput are stored in sWeight. Application specific weights for accuracy, progress, latency and execution time are stored in aWeight. Using these, the system performance and application performance of running application\-j on node\-i would be calculated. Using these metrics, we would be able to calculate pVal which combines both application and system performance into single metric. The applications which need their updates immediately would be classified as green(priority one), next prioritized updates would be yellow (priority two), updates with least priority would be blue (priority three) and classifiers which need not be updated are red. Green, yellow, blue and red are the coloring scheme maintained by coalescer in order to assign priority to an application's model update (refer to Figure~\ref{fig:assprior}).
In the DSOC approach, if individual applications are consistent and make efficient progress towards end goal by carefully considering model updates, we can state that the overall system performance and progress would be prolific.

\begin{figure}
    \centering
    \includegraphics[width=8cm]{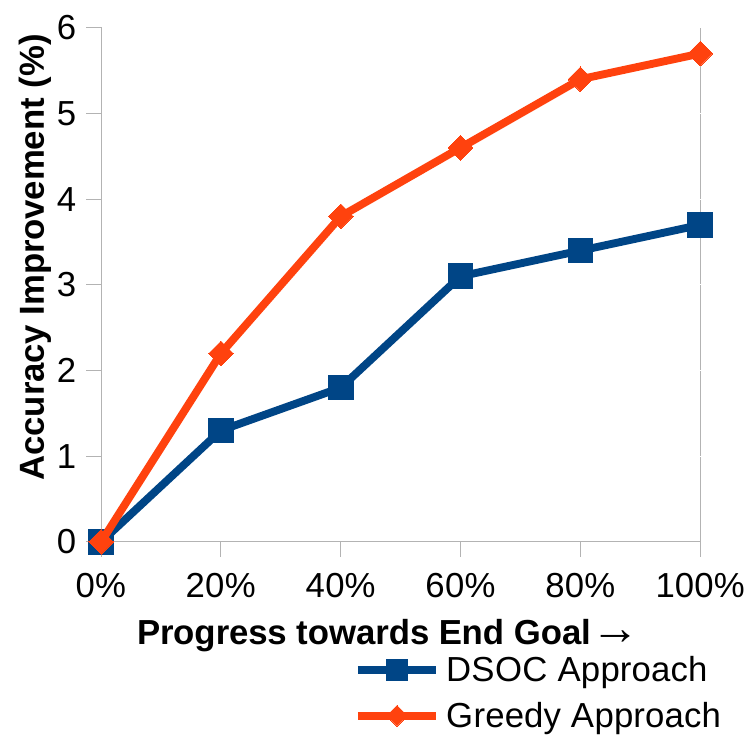}
    \caption{Accuracy improvement against task completion}
    \label{fig:implement2pred}
\end{figure}

In Figure~\ref{fig:implement2pred}, we try to carefully predict the trade-off between accuracy improvement and closeness towards end-goal. Closeness towards end-goal is percentage of task completed like 20\%, 40\%, 60\% and so on. In Greedy approach, the accuracy constantly increases and we reach faster towards the ends goal. We reach the end goal with slightly better accuracy using greedy approach using a lot of resources and performing many updates. On the other hand if we choose DSOC approach, there would be slight improvement in overall accuracy as we progress towards end-goal and there would be slower progress towards the end goal, but it uses less resources and performs fewer updates. In DSOC approach, we reach the end goal with lesser updates and slightly lesser accuracy compared to greedy model.  
%Planning to remove this graph... SO commented..
%\begin{figure}
%    \centering
%    \includegraphics[width=8cm]{figures/implement_1_pred.pdf}
%    \caption{Throughput versus \# of applications}
%    \label{fig:implement1pred}
%\end{figure}

%We need to understand how throughput varies as we scale the number of applications used.  From Figure~\ref{fig:implement1pred}, for greedy approach as we scale the number of applications the throughput increases. As more and more applications are added, many models need to be updated for accuracy improvement. Whenever there's a newer model which gives slight improvement in accuracy, Coalescer updates the application's model. This causes a sharp elevation in throughput. In our proposed xxx approach, not all the models are updated. The coalescer judiciously considers performance metrics and updates only few models with highest priority. When newer models are available which improves an application's accuracy, Coalescer considers numerous factors and prioritizes models to be  updated. Then Coalescer selects a few highest priority models which need to be immediately updated. There would be little increase in throughput as compared to the greedy approach.  

\section{Related Work}
To the best of our knowledge, so far no approaches with focus of this paper (Efficient Patching for coordinating Edge Applications) have been published. In this section, we discuss related work which are closely related to the research problem of this paper. Lele Ma et al.~\cite{service_handoff} proposed efficient service hand-off across edge servers using Docker containers migration. Researchers give an in-depth explanation of leveraging Docker features to the full extent. The paper incorporates migration algorithm for service hand-off which gives insights on the process of patching an application. Taherizadeh et al.~\cite{dml_scaling} proposed an auto-scaling method for time-critical cloud applications considering system performance and application level monitoring. The researchers built a Dynamic Multi-level autoscaling system using Kubernetes as an orchestrator. Kaewkasi et al.~\cite{7886112} worked on building Ant colony optimization based scheduling algorithm which outperforms the built-in scheduling provided by Docker Swarm. This research gave hints on carefully considering resource utilization and available resources for coordinating applications.

%1. Efficient service hand-off for VM transfer across edge devices
%2. Dynamic Multi-level Auto-scaling Rules for Containerized Applications
%3. (Possibly) Improvement of Container Scheduling for Docker using Ant Colony Optimization
\section{Conclusion and future work}

%%%%%%% REWORK, Don't go into specifics ... Make this a point. %%%%%%
Autonomous systems are evolving at a very fast pace and moving towards achieving full autonomy~\cite{transportresearch}. The industry and research community need to focus on coordinating multiple applications that work closely together towards achieving an end goal. Containers are increasing in popularity for building and shipping applications efficiently~\cite{docker_article}. This paper is an early work which focuses on building an orchestrator component using Docker Swarm mode to coordinate multiple applications that are working together towards an end goal. The orchestrator component is responsible for tracking and choosing the model updates leading to performance improvement. The updates would be prioritized considering system performance and individual application performance. Currently, we have a framework and infrastructure setup to deploy, track and update an application. The future plan is to build several different applications using different Machine Learning techniques. We plan to build applications such as intruder detection, simple face recognition, obstacle detection, mission planner which can work collectively towards safely reaching the destination from a source point. During the mission, we try to run different workload by constraining the mission to measure performance of system and record how DSOC efficiently reaches the end goal.

{\small \noindent {\bf Acknowledgments:} This work was funded in part by NSF Grants 1749501 and 1350941 with support from NSF CENTRA collaborations (grant 1550126). This was an IDEA and Early work paper submitted to ICAC 2019 (now known as ASCOS).}

%\- Talk about current approach and algorithm.
%\ - Talk about future work and experiments to be conducted.
%Autonomous Driving vehicles are evolving at a very fast pace. Service hand-off through application migration is a crucial research area in autonomous vehicles. In this paper, we present Chimera Engine and cloudlet which aid in balancing the trade-off between migration time, accuracy, storage and energy requirements while running a Machine Learning application. Our proposed solution is capable of dynamically adapting to field data with Machine Learning models that sufficiently meet the autonomy criteria. This work is important when 1. Multiple applications need to be context switched using an edge device 2. Handling multiple autonomous systems (called swarms) that are coordinating and directing towards a common goal.

%%%%%%%%%%%%%%%%%%%%%%%%%%%%%%%%%%%%%%%%%%%%%%%%%%%%%%%%%%%%%%%%%%%%%%%%%%%%%%%%

{\small
\bibliographystyle{plain}
\bibliography{bibliography}
}

\end{document}